\documentclass[prb,twocolumn,superscriptaddress,showpacs,amsmath,amssymb]{revtex4}
\usepackage{amssymb}
\usepackage{amsmath}
\usepackage{graphicx}
\usepackage{bm}
\begin{document}
\title{Glassy magnetic behavior induced by $Cu^{2+}$ substitution in frustrated antiferromagnet $ZnCr_2O_4$}
\author{Li-qin Yan}
\email{lqyan@aphy.iphy.ac.cn}
\affiliation{National Laboratory for Condensed Matter Physics,Institute of Physics, Chinese Academy of Sciences, Beijing 100080,China}
\author{Ferran Maci\'a}
\affiliation{Departament de F¨ªsica Fonamental, Facultat de F¨ªsica, Universitat de Barcelona, Avda. Diagonal 647, Planta 4, Edifici nou, 08028 Barcelona, Spain}
\author{Jun-rong Zhang}
\author{Zhong-wei Jiang}
\author{Jun Shen}
\author{Lun-hua He}
\author{Fang-wei Wang}
\affiliation{National Laboratory for Condensed Matter Physics,Institute of Physics, Chinese Academy of Sciences, Beijing 100080,China}

\date{\today}

\begin{abstract}
Structure and magnetic properties of the compounds $Zn_{1-x}Cu_xCr_2O_4$ (ZCCO) are investigated systematically. A structural phase transition from space-group symmetry $Fd3m$ to $I4_1/amd$ is observed in ZCCO. The critical value of the doping, $x$, appears at $0.58\sim 0.62$ through the appearance of a splitting of diffraction peaks at room temperature. Measurements of dc magnetization, ac susceptibility, memory effect and exchange bias-like (EB-like) effect have been performed to reveal the glassy magnetic behaviors of ZCCO. The system with $x\leqslant 0.50$ is suggested as a spin glass-like (SG-like) of magnetic characterization whereas doping values of $0.58\leqslant x\leqslant 0.90$ defines the system as a $``$cluster glass-like$"$ (CG-like) with unidirectional anisotropy. The Cu content suppresses the geometrical frustration of $ZnCr_2O_4$, which may correlate with the pinning effect of Cu sublattice on Cr sublattice to a preferential direction.
\end{abstract}

\maketitle

\section{introduction}
The $``$magnetic geometrical frustration$"$ where lattice geometry results in frustration of the antiferromagnetic (AFM) exchange interaction1 is one of the exciting topics of research in modern condensed matter physics\cite{1}. It has been largely studied and observed in system of spinels, pyrochlores and kagome lattices\cite{2,3,4}. Among these systems, the AFM chromium spinel oxides $AB_2O_4$, composed by $AO_4$ tetrahedrons and $BO_6$ octahedrons, have attracted a considerable interest\cite{5,6,7,8}. One example is the case of $ZnCr_2O_4$,\cite{5,6,7,8,9,10,11}where Zn-site ions form a diamond lattice and a strongly geometrical frustration (frustration factor $\approx 31$, defined by $f=|\Theta _{CW}/T_N|)$ dominates spins residing on octahedral Cr-sites with half-filled $t_{2g}$ orbitals, forming a lattice equivalent to the pyrochlore structure. With cooling, this magnet undergoes a spin-Peierls-like phase transition $(T_N=12.5$ K) with AFM order and a tetragonal structural distortion, together with a relief of the geometrical frustration.
 
The spin-glass(SG) property in geometrical frustrated magnets has also been extensively studied for many years\cite{12,13,14,15,16,17,18,19}. It has shown that, experimentally and theoretically,  the ground state degeneracy in geometrical frustration can be removed by atomic disorder or bond disorder leading to a SG type of ordering\cite{17,20,21,22}. For example, $ZnCr_2O_4$ was investigated by A-site small substitution of nonmagnetic ions Cd. This system presents a SG transition instead of AFM transition at $T_N$. The authors proposed that the small Cd doping was to modify the Cr-Cr interaction in a random fashion, leading to bond disorder. It is known that the presence of the disorder coming from either the site disorder or the competing interaction between AFM and ferromagnetism (FM), often generates a conventional SG (CSG) state. However, in systems which possess disorder and highly geometric frustration often display some unconventional SG behavior, usually named as $``$geometrical SG(GSG)$"$. 
A recent study on magnetic-ions $Cu^{2+}$ substitution for $Cd_{2+}$ ions in $CdCr_2O_4$ series (CCCO) has revealed the structural phase transition from cube to tetrahedron at $x=0.64$\cite{24}. $Cd_{0.5}Cu_{0.5}Cr_2O_4$ shows a transition induced by magnetic field from a nonequilibrium SG state to a possible AFM/FM phase separation\cite{23}. At the same tiem, magnetic ions substitution for non-magnetic ions is proposed as a feasible method to partially destroy the magnetic frustration through the pinning effect of the magnetic ions on the magnetic frustrated sublattice\cite{2,23,25}. This pinning effect would establish a preferential orientation within the frustrated sublattice.

However, extensive studies on other compounds of this family are still lacking, and it is of interest to intensify studies in this direction; particularly, magnetic properties of signatures of tetrahedral structures (such as higher Cu substitution for Cd) have been rarely reported. In the present article, we have subjected $Zn_{1-x}Cu_xCr_2O_4$ (ZCCO) to detailed study of structure and magnetic properties by dc and ac magnetic measurements, isothermal magnetization ($M$), memory effect and exchanged bias-like effect.A structure phase transition and an evolvement of magnetic glassy behavior with Cu concentration are revealed. 

The article is organized as follows. In the next section (Sec.II), we detailed the characteristics of the samples as well as the parameters and equipments used in all different measurements. In Sec.III we show the structure phase transition and corresponding analysis via Rietveld refinement (A), the dc magnetic characterization (B), the studies of the ac susceptibility of the samples $Zn_{0.7}Cu_{0.3}Cr_2O_4$, $Zn_{0.5}Cu_{0.5}Cr_2O_4$ and $Zn_{0.1}Cu_{0.9}Cr_2O_4$ (C), memory effect in the sample of $Zn_{0.5}Cu_{0.5}Cr_2O_4$ (D) and EB-like effect of $Zn_{0.1}Cu_{0.9}Cr_2O_4$. In Sec. IV we discuss the results exposed and, finally, we summarize and conclude our study in Sec.V. 

\section{experiment}

\subsection{Sample preparation}
The polycrystalline samples of ZCCO with $x$=0, 0.1, 0.3, 0.5, 0.58, 0.60, 0.62, 0.64, 0.66, 0.68, 0.70, 0.90 have been prepared by using standard solid-state reactions of ZnO ($99.99\%$), CuO ($99.995\%$), and $Cr_2O_3 (99.98\%)$ powders from Alfa Aesar company as starting materials. Once the stoichiometric powder mixtures were well ground, the mixture was heated according to the following heat treatment in air: $4^{\circ} C/min$ up to $200^{\circ}C$ holding 30 min and $4^{\circ}C/min$ up to $830^{\circ}C$ holding 240 min. The resultant mixtures were reground, palletized, and then sintered during 4320 min at $950^{\circ}C$ with subsequent slow cooling to room temperature. Then the compounds were reground and reheated at $980^{\circ}C$ for 7200 min. A black product with a spinel structure was obtained. Powder x-ray diffraction (XRD) pattern confirmed that the final product has a cubic-type spinel single phase for $0\leqslant x\leqslant 0.5$ and tetragonal-type spinel single phase for $0.58\leqslant x\leqslant 0.9$. 

\subsection{Structural analysis}
X-ray powder diffraction experiments were performed by using a Rigaku diffractometer equipped with a monochromator ($CuK\alpha$ radiation, 40 kV, 120 mA). Diffraction data were collected at room temperature in the angle $(2\theta)$ range from $10^{\circ}$ to $80^{\circ}$ with a step of $0.02^{\circ}$. The counting time was 2 s for each step. Crystallographic parameters were analyzed by Rietveld full-profile refinement using the Fullprof program\cite{26}.

\subsection{Magnetic measurements}
The temperature and field dependence of the magnetization were measured by using a commercial superconducting quantum interference device magnetometer (MPMS-7, Quantum Design). For the dc magnetization-temperature curves, measurements were performed after zero-field-cooling (ZFC) and field-cooling (FC) in a fixed field (500 Oe). The magnetization curves for all samples were measured at 5 K and in the magnetic field range from 0 to 5 T, 5 T to -5 T, then -5 T to 5 T to detect the coercive field. For clarify, we showed the curves only from 0 to 5 T in the paper.Ac susceptibility measurements were done using a Physical Property Measurement System (Quantum Design). The data were collected after ZFC. Memory effect and EB-like effect measurements were also performed by MPMS-7, Quantum Design. 

\section{results}
\begin{figure}
\includegraphics[width=0.45\textwidth]{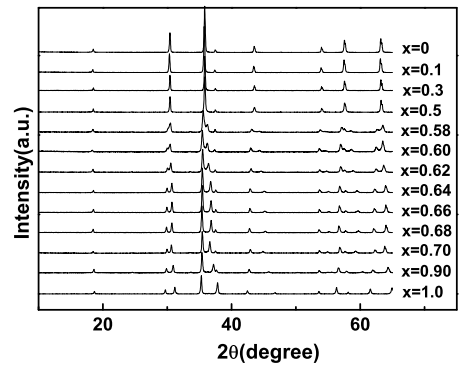}
\label{fig1}
\caption{XRD patterns for all ZCCO compounds.}
\end{figure}
\subsection{Crystal structures}
For clarify, figure 1 shows the powder XRD patterns only within the range from $2\theta = 25^{\circ}$ to $2\theta = 65^{\circ}$ for all the studied ZCCO compounds. It is clear that the diffraction single peaks of $30^{\circ}$, $36^{\circ}$, $43^{\circ}$,$54^{\circ}$,$58^{\circ}$ and $63^{\circ}$ split into double peaks with increasing the copper dopant, which indicates that a structure phase transition has occurred. To determine accurate lattice parameters and atomic positions, Rietveld analysis was carried out for all the powder XRD data. Structure phase with space group $Fd3m$ and the initial sets of Zn(Cu) at 8a(1/8,1/8,1/8), Cr at 16d(1/2,1/2,1/2) and O at 32e($x,x,x$) changes into a new phase with space group $I4_1/amd$ and sets of Zn(Cu) at 4a(0,1/4,7/8), Cr at 8d(0,1/2,1/2) and O at 16e(0,$y$,$z$) within the range $0.58\leqslant x\leqslant 0.62$. Notice that our previous study in $Cd_{1-x}Cu_xCr_2O_4$(CCCO) series\cite{24} showed that CCCO underwent a phase transformation from cubic $(Fd3m)$ to tetragonal $(I\overset{-}42d)$ at Cu concentration of $x$ = 0.64. This distortion was associated with the big difference in the ionic radii ($Cd^{2+}: 0.97\mathring A$, $Cu^{2+}: 0.72\mathring{A}$) and Jahn-Teller effect of $Cu^{2+}$ ions\cite{27}. Compared with the noncentrosymmetric $(I\overset{-}42d)$ of the CCCO series\cite{24}, the phase transition to $I4_1/amd$ in ZCCO is more centrosymmetric (two less alterable coordination parameters: $X_{Cr}$ and $X_O$), attributable to the smaller difference of ionic radii on A-sites of $Zn^{2+} (0.74\mathring{A})$ compounds compared with $Cd^{2+} (0.97\mathring{A})$ ones.
The results of the lattice parameters and unit cell volume are plotted in Fig. 2 as a function of the Cu concentration. An abrupt drop in volume cell [see upper inset of Fig.2] is observed at $x$ = 0.58, which is a result of the smaller size of the ionic radius of $Cu^{2+} (0.72\mathring{A})$ compared with $Zn^{2+} (0.74\mathring{A})$. This contraction corresponds to the compression of the $c$ axis and a slight expansion of the $a$and $b$ axes [see Fig. 2], indicating that the ZCCO structure is distorted from cube to tetrahedron at $x$ = 0.58. The distorted feature, $c/a$, first increases abruptly, then decreases with the increasing Cu content[see lower inset of Fig.2], indicating that the most distorted case occurs at $x$ = 0.58 with $c/a$ = 1.385. 
\begin{figure}
\includegraphics[width=0.5\textwidth]{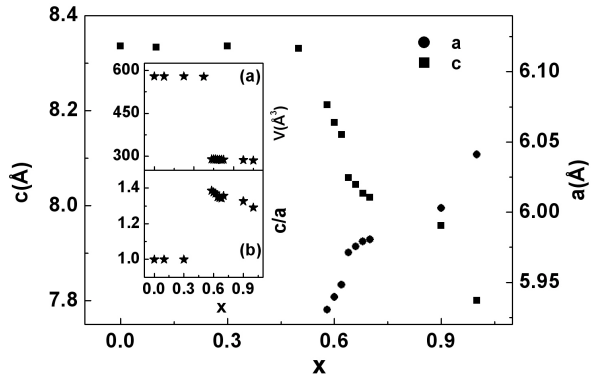}
\label{fig2}
\caption{$X$ dependence of crystal lattice parameters. The insets display the $x$ dependence of (a) volumn and (b) $c/a$.}
\end{figure}
\subsection{Magnetization}
Figure 3 presents the dc magnetic measurements for different Cu content $x$=0, 0.1, 0.3, 0.5, 0.58, 0.7, 0.9 of the $M_{ZFC}(T)$ and $M_{FC}(T)$  under an applied field of 500 Oe. When the sample is doped with $0.58\leqslant x\leqslant 0.9$, going from high to low temperatures, the $M_{ZFC}(T)$ increases till a maximum and then drops down with a small or obvious change in the downward trend below $\sim$50 K until the lowest temperature, $T$ = 5 K. The $M_{ZFC}(T)$ and $M_{FC}(T)$ have the same values for temperatures above $T_a$ (defined as the bifurcation temperature between ZFC and FC curves). Below this temperature, $M_{FC}(T)$ rises more rapidly until a certain point from where it still have an slowly upward trend until 5 K. $T_a$ shifts to higher temperatures with higher $x$. Such a large bifurcation between ZFC and FC curves is a hint of an existence of a short-ranged-ordered cluster, as observed in some perovskites\cite{28,29,30}. When the samples are doped with $0\leqslant x\leqslant 0.5$, with decreasing the Cu content, the difference between cusp temperature and $T_a$ in $M_{ZFC}(T)$ curves is reduced and even coincides at $x$=0, indicating now the reduction of the amount of the short-ranged-ordered clusters and the emergence of SG-like behavior. 
\begin{figure}
\includegraphics[width=0.40\textwidth]{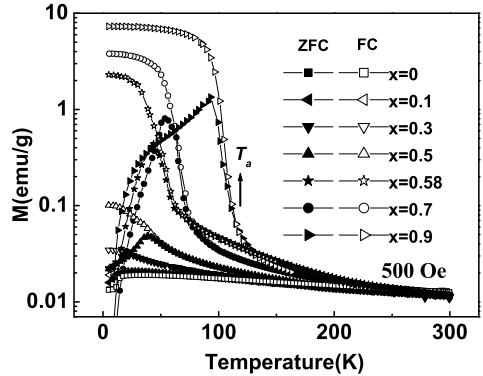}
\label{fig3}
\caption{Temperature dependence of ZFC and FC magnetization for ZCCO compounds under the magnetic field of 500 Oe.}
\end{figure}

The magnetization data above 200 K can be well fitted, for all samples, using the Curie-Weiss law with the expression $\chi^{-1}= (T-\theta)/C$, where $\theta$ denotes the Weiss constant and $C$ is the Curie constant. Figure 4 shows the corresponding fitting parameters. One can see that both the Curie-Weiss temperature and the N\'eel temperature (defined as the maximum value of $|dM/dT|$) increase with $x$, suggesting an increase of the ferromagnetic interaction contributed from Cu sublattice. It is known that the spinel $CuCr_2O_4$,\cite{27}with a tetrahedral crystal structure and  quenched magnetic frustration at room temperature, in which the magnetic moment of the Cu sublattice is FM while the interaction with the Cr sublattice becomes Ferrimagnetic (FI). On the other hand, the negative Weiss constant is indicative of the antiferromagnetic correlations dominance from the Cr sublattice. In order to analyze the geometrical frustration, we also plotted the frustrated factor $f=|T_{CW}/T_N|$ vs. $x$ in Fig. 4. It is then found that the strong frustration of $ZnCr_2O_4$ turns out to be suppressed with Cu content, which has also been attained in CCCO system and could be ascribed to the pinning effect of Cu sublattice on the geometrical frustrated spins. Based on the formula $\mu_{eff} =(3k_{B}c/N_A)^{1/2}$ ($k_B$ and $N_A$ being the Boltzmann constant and the Avogadro number), the effective magnetic moment $\mu_{eff}$ can also be determined [see inset of Fig.4]. The value of $\mu_{eff}$ decreases with Cu content while the theoretical value, according to the formula $2[s(s+1)]1/2\mu_B$, (3.87$\mu_B$ and 1.73$\mu_B$ for $Cr^{3+}$ and $Cu^{2+}$ ions with spin s=3/2 and s=1/2, respectively) slightly increases. The experimental values of $\mu_{eff}$ are lower than the theoretical evaluation, which should orignate from the short-range interaction in paramagnetic region. Additionally, the trend discrepancy between the measured and theoretical values implies an enhanced magnetic correlation with Cu content above the N\'eel temperature and a formation of step grown FI clusters.
\begin{figure}
\includegraphics[width=0.40\textwidth]{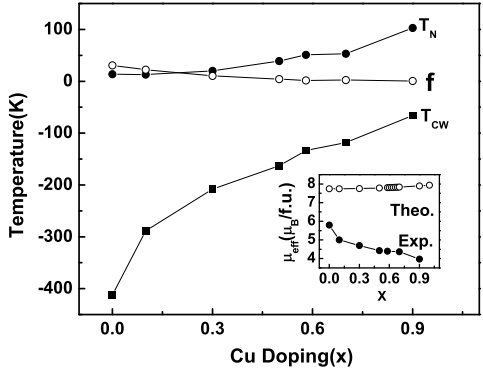}
\label{fig4}
\caption{$\protect{X}$ dependence of Curie temperature $T_N$, Curie-Weiss temperature $T_{CW}$ and frustration factor $f$. The inset shows the $x$ dependence of effective magnetic moment per formula theoretically and experimentally.}
\end{figure}

Figure 5(a) shows the isothermal magnetization measurements at 5 K for all samples. For the samples with $x\leqslant 0.5$, $M$ varies nearly linearly and nonhysteretically with $H$, characteristic of dominance by antiferromagnets. For $x\geqslant 0.58$, a curvature appears beyond $\sim$ 1 T together with an observable S-type behavior. The hysteretic behavior and the absent saturation tendency for the samples with $0.58\leqslant x\leqslant 0.90$ imply a coexistence of AFM and FM coupling, thereby pointing towards SG or cluster glass freezing. The magnetization values at 5 T increases with $x$ at a rate of $0.1\mu_B$ per $Cu^{2+}$ ion for the samples with $0.1\geqslant x\geqslant 0.5$, smaller than the expected saturation moment, $1\mu_B$, implying an absence of long-range-ordered ferromagnetic Cu sublattice. However, for the samples with $0.58\leqslant x\leqslant 0.90$[see Fig. 5(b)], $1.3\mu_B$ per $Cu^{2+}$ ion is obtained, which is larger than the expected saturation moment, $1\mu_B$, implying an existence of a parallel addictive magnetic contribution to $Cu^{2+}$ ferromagnetic sublattice. A metamagnetic-like transition is observed at a critical field $H_{cr}$, (defined as the maximum value of $|dM/dH|$) [See Fig. 5(c)]. The value of $H_{cr}$ is about 1.24 T at $x$=0.58. Both the $H_{cr}$ and coercive field $H_c$ [See the inset of Fig.5(c) and Fig.5(d)] decrease with Cu content, which is in accordance with the released structural distortion with $x$ in tetrahedral spinels. The magnetic anisotropy appears at $x$=0.58 and it is reduced with further increasing Cu content, having the maximum value at $x$=0.58.
The increasing A-site substitution results in the following effects: (i) The magnetic frustration is continuously suppressed with the Cu content, resulting from an increase of $T_{N}$ and a decrease of $|\Theta_{CW}|$. (ii) The magnetization of the samples increase significantly with $x$. For tetrahedral-type spinels of ZCCO, an addictive moment of 0.3$\mu_{B}/f.u.$ is obtained with $x$, implying a pinning interaction of Cu sublattice to Cr frustrated magnetic lattice; (ii) The system is proposed to transform from AFM to SG-like state then to short-range-ordered cluster state with $x$; (iii) In tetrahedral-type ZCCO spinels of $0.58\leqslant x\leqslant 0.90$, both the metamagnetic-like transiton field $H_{cr}$ and coercive field $H_c$ are decreased with $x$, indicating the increasing Cu content recudes the magnetic anisotropy. 
\begin{figure}
\includegraphics[width=0.50\textwidth]{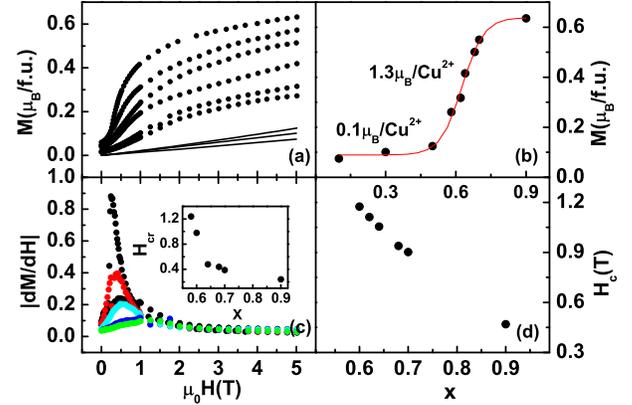}
\label{fig5}
\caption{(Color online) (a)The magnetic field dependence of magnetization for ZCCO compounds at 5 K.(b)The accordingly Cu content dependence of magnetic moment per formula at 5 T and 5 K.(c) The magnetic field dependence of $|dM/dH|$. The inset shows the accordingly Cu content dependence of metamagnetic transition field $H_{cr}$ at 5 K.(d) The Cu content dependence of coercive field of $H_{c}$ for ZCCO compounds at 5 K.}
\end{figure} 
\subsection{Ac susceptibility }
In order to investigate the glassy magnetic property in ZCCO, ac susceptibility measurements, such as frequency and magnetic field dependence of transition temperature, are necessary. In this work, only the cubic spinels $Zn_{0.7}Cu_{0.3}Cr_2O_4$, $Zn_{0.5}Cu_{0.5}Cr_2O_4$ and tetragonal spinel $Zn_{0.1}Cu_{0.9}Cr_2O_4$ are discussed as examples. In Fig. 6 is presented the temperature dependence of the ZFC in-phase $\chi^{'}(T)$ at different frequencies and an ac field of 5 Oe for $Zn_{0.7}Cu_{0.3}Cr_2O_4$. The peak amplitude decreases with (increasing) frequency. However, the peak temperature is frequency independent, indicating the predominance of AFM long-range ordering and the deficiency of conventional SG-like state. Fig. 7(a) presents the temperature dependence of the ZFC in-phase $\chi^{'}(T)$ at different frequencies and with an ac field of 5 Oe for $Zn_{0.5}Cu_{0.5}Cr_2O_4$. The $\chi^{'}(T)$ curves display a peak at $T_f(\omega)$, which is frequency dependent, having the values 38, 38.10, 38.24, 38.35 and 38.39K for $\omega/2\pi$ = 333, 999, 3333, 7777 and 9999 Hz, respectively. The temperature of the peak, $T_f(\omega)$, shifts towards higher temperatures and the peak amplitude diminishes with increasing frequency. The value of the frequency sensitivity of $T_f(\omega)$, $\Delta T_f(\omega)/[T_f(\omega)\Delta log_{10}\omega]$, has been a criterion for the presence of a canonical SG from SG-like\cite{28}. It is about 0.007 for $\chi^{'}(\omega,T)$, lower than those reported for other typical insulating SG systems\cite{31}. The divergence of the maximum relaxation time $\tau_{max}$, occurring at the peak temperature, can be investigated by using conventional critical slowing down:
\begin{equation}
\frac{\tau}{\tau_0}=\xi^{-z\nu}=\bigglb(\frac{T_f(\omega)-T_f}{T_f}\biggrb)^{-z\nu}
\end{equation}
A best fit of the measured data to Eq. (1) is shown in the inset of Fig. 7(a), yielding the values $\tau_0=2.77\times 10^{-13}s$ , $z\nu$=4.55 and the transition temperature $T_f$ = 37.65 K. Both values of $\tau_0$ and $z\nu$ are the typical values for conventional spin glasses\cite{32,33}. However, $\tau_0=7.7\times 10^{-10}s$ and $T_f =36.86$ K are obtained by fitting to the Volgel-Fulcher scaling law (Eq. 2),
 \begin{equation}
\frac{\tau}{\tau_0}\propto \exp{\frac{E_a}{k(T_f(\omega)-T_f)}}
\end{equation} 
The different dynamic scaling indicates, therefore, that there is a divergence of the SG relaxation time at a finite transition temperature, which demonstrates a phase transition from paramagnetism (PM) to SG-like combined with some small AFM clusters. Also the existence of small AFM clusters can be confirmed by the shape of ac-susceptibility peaks, which is not very sharp. Furthermore, it is well known that $T_f$ shifts to lower temperature with the applied dc field in many classical SG systems, which can be well described by the Almeida-Thouless (AT) line\cite{34}.
 \begin{equation}
{T_f(H)}=1-bH^\delta
\end{equation} 
The temperature dependence of the ac susceptibility (in the in-phase component) under dc fields of 0, 500, 2000 Oe measured at a frequency of 333 Hz is displayed in Fig.7(b). Both the peak amplitude and the peak temperature decrease with increasing applied field. The inset of Fig.7(b) plots the experimental $T_f(H)$ values and the corresponding fitted data to Eq.(3). The fitted value of the exponent $\delta$ is 0.898, slightly larger than the value $\delta=2/3$ given by the mean-field theory prediction for SG\cite{35,36}. However, this is only a necessary (but not a sufficient) feature of a SG transition\cite{36}. 
\begin{figure}
\includegraphics[width=0.40\textwidth]{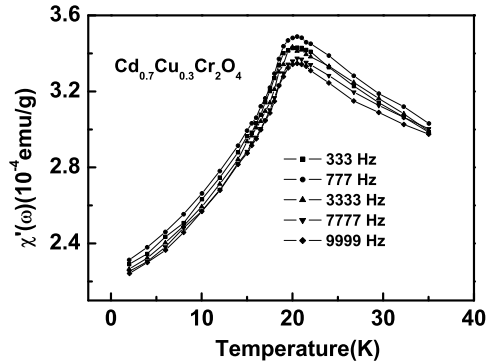}
\label{fig6}
\caption{Temperature dependence of $\chi^{'}(\omega,T)$ for $\omega /2\pi$ =333,777,3333,7777,9999 Hz for $Zn_{0.7}Cu_{0.3}Cr_2O_4$.}
\end{figure}
For $Zn_{0.1}Cu_{0.9}Cr_2O_4$, as shown in Fig.8, with decreasing temperature, the susceptibilities abruptly increase until the maximum at 90 K showing at $T_N =104$ K a small change in the upward trend, there is no frequency dependence for the peak of $\chi^{'}(T)$ at 90 K, exhibiting the PM-FI phase transition. With decreasing temperature, the CG-like behavior is exposed by a shoulder in $\chi^{'}(T)$ at $T_{f}^{'}\approx 50 $ K and a hump in $\chi^"(T)$ at $T_f^"$ as marked by the arrow in Fig. 8. $T_{f}^{'}$ and $T_{f}"$ are expected to be frequency dependent because of the CG-like freezing. The exact frequency dependence position of the shoulder in $\chi^{'}(T)$, $T_{f}^{'}$, is rather difficult to determine. On the other hand, $T_{f}^"$ shifts considerably to lower temperatures with decreasing frequency, with values $T_{f}^"\approx$ 47.7, 50.3, 51.1, 54.3 and 56.1 K for $\omega/2\pi$ =333, 999, 3333, 7777 and 9999 Hz, respectively. The sensitivity of $T_f(\omega)$,$\Delta T_f(\omega)/[T_f(\omega)\Delta log_{10}\omega]$is about 0.119 for $\chi^"(\omega,T)$, being between the reported insulating SG, $(FeMg)Cl_2$, and superparamagnet, $a-(Ho_2O_3)(B_2O_3)$, indicating a CG-like freezing\cite{31}. An attempt to fitting $T_{f}"(\omega/2\pi)$ data using Eq.(1), gives $\tau_{0}=2.11\times 10^{-7}s$ and $z\nu =5.979$ when $T_f = 37.3$ K is taken by fitting to the Volgel-Fulcher scaling law Eq.(2). The large relaxation time implies a freezing process of large size cluster glass at $T_{f}^"$.
\begin{figure}
\includegraphics[width=0.35\textwidth]{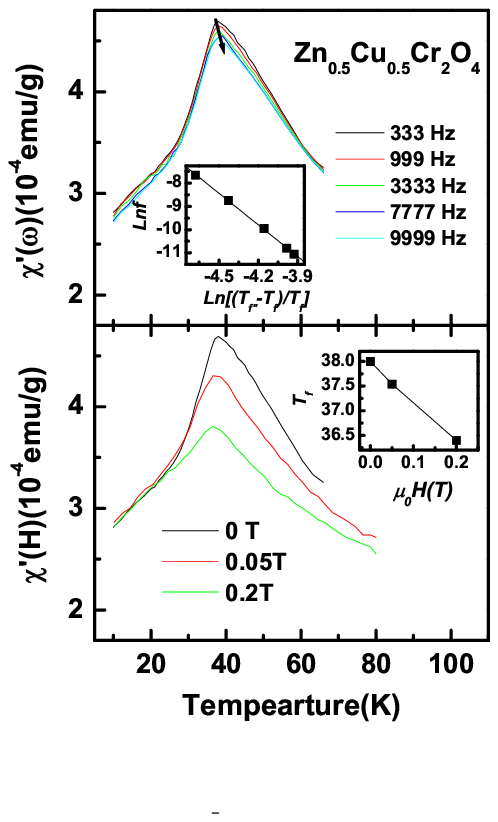}
\label{fig7}
\caption{(Color online) (a)Temperature dependence of $\chi^{'}(\omega,T)$ for $\omega /2\pi$=333,999,3333,7777,9999 Hz for $Zn_{0.5}Cu_{0.5}Cr_2O_4$ compound. The inset shows the measured freezing temperature $T_{f}(\omega,T)$ and the best fitted line by Eq.(1) for $\chi^{'}$. (b)Temperature dependence of $\chi^{'}(H,T)$ measured at a frequency of 333 Hz under the magnetic fields of 0, 500 and 2000 Oe. The inset displays the experimental $T_{f}(H)$ values and the fitted data to Eq.(2).}  
\end{figure}

The above ac susceptibility results show that the cubic spinel of ZCCO presents a SG-like behavior while the tetrahedral one exhibits a coexistence of FI and CG-like behavior. 
\begin{figure}
\includegraphics[width=0.60\textwidth]{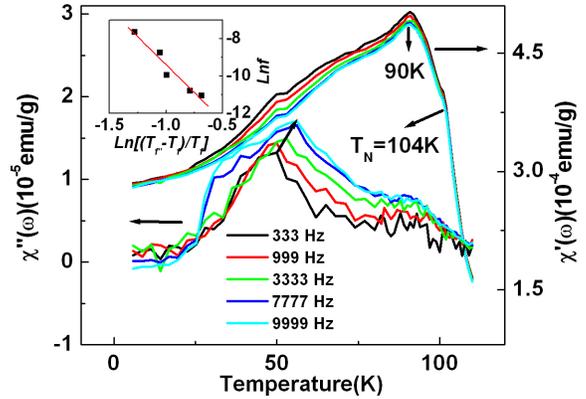}
\label{fig8}
\caption{(Color online) Temperature dependence of in-phase $\chi^{'}(\omega,T)$ and out-of-phase $\chi^"(\omega,T)$ ac susceptibilities measured at various frequencies of $\omega /2\pi$=333,999,3333,7777,9999 Hz for $Zn_{0.1}Cu_{0.9}Cr_2O_4$.The inset shows the measured freezing temperature $T_{f}(\omega,T)$ and the best fitted line by Eq.(1) for $\chi^"$.}
\end{figure}
\subsection{Memory effect}
For further probe the nature of the glassy behavior in ZCCO spinels, the time response of dc magnetization is important to reveal the spin dynamics\cite{37}. In this work, we demonstrate the characteristic behaviors of glassiness by the memory effect for $Cd_{0.5}Cu_{0.5}Cr_2O_4$. It was studied by employing a dc magnetization method that was originally developed for the study of interacting glassy system \cite{38,39} In a SG or a system of interacting magnetic particles (but not in noninteracting particles), a dip appears on reheating at the temperature at which the sample was stopped under zero field.

The sample was first zero-field-cooled (ZFC) from 300 K to 5 K continuously at a cooling rate of 2 K/min. After reaching the bottom temperature, a 50 Oe field was applied and the magnetization was measured on heating at the same rate up to 100 K. This $M$-$T$ curve was referred as the reference curve, shown as a solid line in Fig. 9. Then, the sample was cooled again from 300 K to 5 K at the same rate in zero field but with a temporary stop at 28 K for a time $t_w=7 000 s$. Finally, the magnetization in a 50 Oe field was recorded again during heating. The obtained results are shown in Fig. 9. The difference between two $M(T)$ curves, $M=M(T)-M_{ref}(T)$, is also shown in Fig.9. A minimum is observed at the stop temperature 28 K, reflecting the memory effect. Additionally,another dip also appears around the SG-like freezing temperature 38 K.The reason is still unclear. An identical measurement procedure at 30 K were also performed for the sample with $x$=0.9. However, we have not observed any dip in the difference between two $M(T)$ curves, implying an existence of noninteracting superferrimagnetic-like clusters. The behavior of the memory effect for $x$=0.5 but $x$=0.9 suggests an evolvement of glassy behavior from interacting particles to superferrimagnetic-like clusters with Cu content.
\subsection{Exchange bias-like effect}
If the ground state in tetragonal spinel is a true CG with superferrimagnetic clusters, the EB-like effect should be observed\cite{40,41}. For simplicity, only experiments performed on a sample of $Zn_{0.1}Cu_{0.9}Cr_2O_4$ with the low magnetic anisotropy among the tetrahedral ZCCO cases, was discussed, although similar effects were also observed in other tetrahedral ZCCO compositions. 
Figure 10 shows the influence of measuring field on the exchange bias for $Zn_{0.1}Cu_{0.9}Cr_2O_4$. For each measurement, the sample was cooled under a field of 200 Oe from 300 to 5 K, then the hysteresis loops were measured between $\pm 0.02$, $\pm 0.05$, $\pm 0.1$, $\pm 0.2$, $\pm 0.3$, $\pm 0.5$, $\pm 0.7$, $\pm 1$, $\pm 2$, $\pm 3$, $\pm 4$ and $\pm 5$ T. When the measuring field was small ($H\leqslant$1 T), the FC hysteresis loops always shifted to the negative field and positive magnetization, suggesting that an unidirectional anisotropy existed after the field cooling. However, when the measuring field was high enough, $H\geqslant 2$ T, the FC hysteresis loops did not show any shift, i.e. EB-like effect disappears at high magnetic fields. The inset of Fig. 10 shows the measuring field dependence of the exchange bias field, $H_E$, and the magnetization shift, $M_E$. $H_E$ is defined as the middle point between the negative field and the positive field at which the magnetization equals to zero. And $M_E$ is defined as the middle value between the two intersection points of the magnetization with $\mu_0H$ = 0 T. These shifts decrease rapidly with increasing the applied magnetic field and disappear around 2 T. It is known that the EB in the AFM/FM systems only shifts in the direction of the field axis\cite{42}. Thus, this EB-like effect in $Cu_{0.9}Zn_{0.1}Cr_2O_4$ should be derived from freezing effect of the random anisotropy in the CG, rather than the coupling effect at the interfaces of the FM/AFM. The spins of the cluster are aligned to the field upon field cooling.
\begin{figure}
\includegraphics[width=0.50\textwidth]{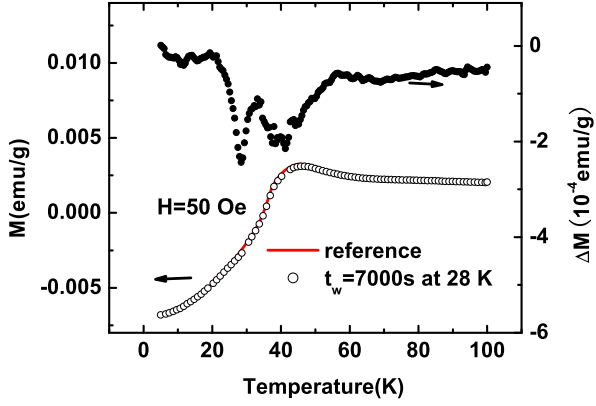}
\label{fig9}
\caption{(Color online) Temperature dependence of the reference magnetization $M_{ref}$ (red solid line), the magnetization $M$(open circle), and $\Delta M=M-M_{ref}$(close circle) for $Cd_{0.5}Cu_{0.5}Cr_2O_4$ compound.}
\end{figure} 
The above EB-like measurements provide somehow an evidence of the freezing of superferrimagnetic cluster with random anisotropy at lower temperatures, forming a CG-like ground state. Random anisotropy of FI cluster causes an EB-like effect for the tetrahedral spinel samples.
\section{DISCUSSION}
The above results confirm that cubic spinels ZCCO ($x\leqslant 0.50$) show a SG-like behavior while tetrahedral spinels ZCCO ($0.58\leqslant x\leqslant 0.90$) exhibit a coexistence of FI and CG-like behavior. In order to understand the physics associated with A-site magnetic ions substitution in geometrical frustrated spinels $ACr_2O_4$, we have constructed a schematic phase diagram of ZCCO[see Fig. 11]. Geometrically frustrated antiferromagnet presents paramagnetic spin liquid (SL) state between $T_N$ and $|T_{CW}|$\cite{43,44}. It would be expected that, for lower Cu concentration below $50\%$, AFM order is dominant, coexisting with geometrical SG (GSG) and conventional SG (CSG) at low temperatures. The GSG component decreases with increasing the Cu content and disappears at $x$=0.5 based on the suppressed frustrated factor. Thereafter FI cluster with glassy characteristics is dominated for the copper concentration above $58\%$ on A-site, exhibiting CG-like behavior below the freezing temperature. Our experiments only prove the existence of SG-like for $x\leqslant 0.50$ and superferrimagnetic-like behavior for the sample with $x$ = 0.90, but do not provide further CG-like evidence for other tetrahedral spinels and any definitive geometrical SG evidence. A detailed experiment measurement and analysis must be performed at the critical region. 
\begin{figure}
\includegraphics[width=4.5\textwidth]{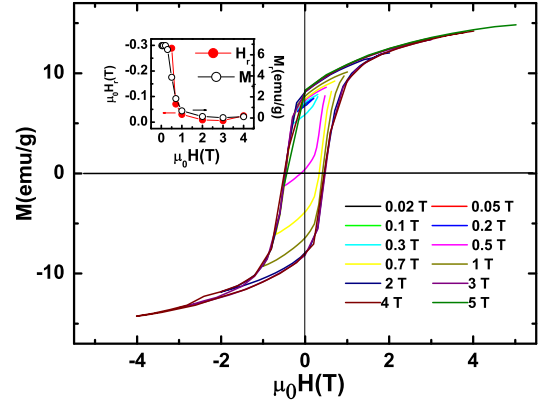}
\label{fig10}
\caption{(Color online) The FC (in 200 Oe) hysteresis loops for $Zn_{0.1}Cu_{0.9}Cr_2O_4$ at 5 K with different measuring magnetic fields. Inset: mesuring field dependence of $H_{E}$ and $M_{E}$ at 5 K.}
\end{figure}

In an ideal ZCCO compound, each Zn/Cu atom is surrounded by four oxygen atoms which build up a tetrahedron. Each Cr atom is surrounded by six oxygen atoms which build up an octahedron. A number of Zn atoms substituted by Cu introduces a magnetic moment that increases not only the crystal distortion but also the magnetic interactions. Here, a phase separation landscape due to the chemical inhomogeneous substitution on A-site is proposed for the series ZCCO by a schematic site percolation model\cite{45,46,47,48,49,50,51,52}. Assuming that a low concentration of $Cu^{2+}$ ions occupies the A-sites in the $ZnCr_2O_4$ crystal lattices, there is an A-site disorder with the appearance of small rich $Cu^{2+}$ clusters and a chemically phase separation emerges. A small quantity of clusters exhibits no interaction between them ascribed to their long separation. Thus, the FM order percentage of Cu sublattice is still low and the long-range AFM order predominates. But local $Cu^{2+}$ ions with short-range-ordered FM have a pinning effect to the surrounded Cr sublattice which induces a local released geometrical frustration, exhibiting a long-range-ordered AFM accompanied with a geometrical SG, as shown in $x$=0.3. When the FM interaction  is comparable to the AFM interaction, large site disorder and frustration appear, and finally, no long-range ordering exists around the intermediate Cu concentration region resulting in a conventional SG-like behavior with a few small AFM clusters, as shown in $x$=0.5. At the nominal copper concentration $x$=0.58, the increase of the A-site substitution makes the random oxygen displacements, originated from ionic radii difference between $Zn^{2+}$ and $Cu^{2+}$ on A-site, become significant. It would enhance the random local radial distortions of the $CrO_6$ octahedron and lead to the reduction of the Cu-Cu, Cu-Cr and Cu-O bonds. To reach the lowest energy state of the system, the structure phase transition occurs from cubic to tetrahedral lattice, where both $Cu/ZnO_4$ and $CrO_6$ polyhedra are distorted due to anisotropic atomic displacement of the oxygen atoms. In the tetrahedral structure, the Cu-Cu interaction is ferromagnetic, which brings about a pinning interaction to the complicated Cr magnetic sublattice. At the same time, the size of FI cluster was expected to grow. The spin flipping time $\tau_0$ is enhanced up to $\tau_0\sim 10^{-7}s$ at $x$=0.9, suggesting the crossover to the glassy state with the FI cluster. The random anisotropy clusters with large grown size and the weak interaction between the clusters appear, exhibiting a collected FI order derived from the intracluster' correlation and a CG-like behavior derived from the interclusters. Also it can be seen that, in the $CuCr_2O_4$ spinel, it is only required about $10\%$ of Zn to suppress the long-range FI order into superferrimagnetic-like state. On the other hand, the $ZnCr_2O_4$ spinel takes $58\%$ of Cu to suppress the AFM order.
\begin{figure}
\includegraphics[width=0.40\textwidth]{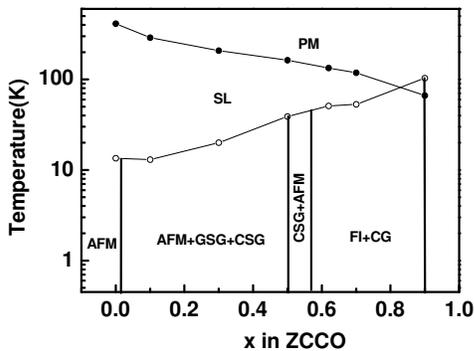}
\label{fig11}
\caption{Doping-temperature $(x-T)$ schematic phase diagram of $Zn_{1-x}Cu_xCr_2O_4$, where $x$ is the nominal Cu concentration.}
\end{figure}

Note that the magnetic properties in $Zn_{0.5}Cu_{0.5}Cr_2O_4$ are different from $Cd_{0.5}Cu_{0.5}Cr_2O_4$. The former shows a conventional SG-like state and lacks any observation of magnetic phase separation induced by applied magnetic field. Considering similarities in the structure between them, it is obvious due to the different ionic radii between $Zn^{2+}$ and $Cd^{2+}$ ions. In $Cd_{0.5}Cu_{0.5}Cr_2O_4$, the larger ionic radii difference on A-site increases the distortion of tetrahedron $AO_4$, resulting in a larger displacement of oxygen coordinates and $Cr^{3+}$ off-center, which finally allows the AFM/FM magnetic phase separation arising from the pinning effect more available\cite{23}. This suggests that the influence of the A-site disorder on the local structure distortion and magnetic behavior can be significant in frustrated spinel magnet. A lower difference between A-site ionic radii, such as ZCCO series, can be expected to be structurally and chemically more homogeneous, reflecting in the structure phase transitions at the different critical concentration of $x_{cu}=0.58$ in ZCCO with $x_{cu}=0.64$ in CCCO, which would reduce probability of magnetic phase separation under external conditions.
\section{CONCLUSIONS}
In conclusion, a series of chromium spinels, $Zn_{1-x}Cu_xCr_2O_4$, with an increasing A-site magnetic ionic substitution from 0.1 to 0.9, focused on $x$=0.5 and 0.9, have been investigated using various experimental techniques. Our results have demonstrated that the samples with $x\lneq 0.58$ have SG-like magnetic behavior with the cubic crystallographic structure and the sample with $x\geqslant 0.58$ have an coexistence of FI and CG-like magnetic behavior with the tetrahedral-type spinel structure. A chemical phase separation scenario is taken to understand these experimental phenomena by a schematic site percolation model.

\acknowledgements{This work was supported by the National Natural Science Foundation of China (Grant No 10505029).}
\bibliography{yan}
\end{document}